\documentclass[a4paper,notoc]{JHEP3}
\usepackage{bbm}
\usepackage{amsmath}
\usepackage{graphicx}

\title{Einstein-aether as a quantum effective field theory}
\author{Benjamin Withers\\Blackett Laboratory, Imperial College London,\\ United Kingdom\\Email: \email{benjamin.withers02@imperial.ac.uk}}

\abstract{
The possibility that Lorentz symmetry is violated in gravitational processes is relatively unconstrained by experiment, in stark contrast with the level of accuracy to which Lorentz symmetry has been confirmed in the matter sector. One model of Lorentz violation in the gravitational sector is Einstein-aether theory, in which Lorentz symmetry is broken by giving a vacuum expectation value to a dynamical vector field. In this paper we analyse the effective theory for quantised gravitational and aether perturbations. We show that this theory possesses a controlled effective expansion within dimensional regularisation, that is, for any process there are a finite number of Feynman diagrams which will contribute to a given order of accuracy. We find that there is no log-running of the two-derivative phenomenological parameters, justifying the use of experimental constraints for these parameters obtained over many orders of magnitude in energy scale. Given the stringent experimental bounds on two-derivative Lorentz-violating operators, we estimate the size of matter Lorentz-violation which arises due to loop effects. This amounts to an estimation of the natural size of coefficients for Lorentz-violating dimension-six matter operators, which in turn can be used to obtain a new bound on the two-derivative parameters of this theory.}

\begin{document}\setlength{\unitlength}{1mm}

%
%
%
%
%
%
%
%
%
%
%
%
%
%
%
%

%
%
%
%
%
%
%
%
%
%
%
%
\section{Introduction}\label{intro}
Lorentz symmetry is widely considered to be one of the cornerstones of modern physics, confirmed to an exquisite level of accuracy by experiment. These investigations are facilitated theoretically by the Standard Model Extension~\cite{Kostelecky:1994rn, Colladay:1996iz,Colladay:1998fq}, which parametrises the renormalisable Lorentz violating operators which can be introduced into the standard model Lagrangian. In this framework~\footnote{See \cite{Coleman:1998ti,Colladay:1996iz,Colladay:1998fq, Mattingly:2008pw, Mattingly:2005re} for a comprehensive review of experimental constraints.}, some Lorentz violating perturbations are excluded at the accuracy of $10^{-27}$.

In light of this experimental evidence it is difficult to imagine that Lorentz symmetry could be violated at experimentally accessible energy scales. Additionally, theoretical difficulties arise even if one considers energy scales which are presently inaccessible. Typical motivations for doing so include possible quantum gravity effects, for example, a Lorentz violating granularity of spacetime near the Planck scale. Recent arguments \cite{Collins:2004bp,Crichigno:2006hp,Collins:2006bw} suggest that effects of Planck scale physics can be communicated to low energy processes via radiative corrections, in particular, Lorentz violating contributions to low dimensional operators are argued to arise from the high-momentum part of loop integrals. Contrary to naive expectations based on the vast separation of scales involved, these contributions are not suppressed by inverse powers of the Planck mass and instead contribute at the percent level. This creates a new fine-tuning problem for these models.

An altogether distinct possibility is that Lorentz symmetry is spontaneously broken at low energies, but is restricted to the gravitational sector~\cite{Kostelecky:1989jp,Kostelecky:1989jw}. In this class of models, Lorentz symmetry is spontaneously broken by a vector field which is given a vacuum expectation value (VEV) using either a potential or a Lagrange multiplier constraint, examples of which include ~\cite{Kostelecky:1988zi,Kostelecky:2003fs,Kostelecky:1991ak,Bluhm:2004ep,Jacobson:2000xp, Eling:2004dk, Zlosnik:2006zu, Gripaios:2004ms,Zhao:2007ce,Kanno:2006ty}.

In this paper we consider a diffeomorphism invariant model, Einstein-aether theory~\cite{Jacobson:2000xp, Eling:2004dk}. In this model, Lorentz symmetry is spontaneously broken by a dynamical vector field known as the aether field, $\tilde{A}$, with a timelike VEV, $v$, enforced through a Lagrange multiplier constraint, $\lambda$. The Einstein-aether Lagrangian, in the mostly minus signature convention $(+,-,-,-)$, and in units where $\hbar=c=1$,
\begin{equation}\label{bare}
S=\int{d^4x \sqrt{-g}\left\{\mathcal{L}_{2}+\mathcal{L}_m+\lambda\left(\tilde{A}^2-v^2\right)\right\}},
\end{equation}
where $\mathcal{L}_m$ represents the matter content and
\begin{eqnarray}\label{l2def}
\mathcal{L}_2 &=& -\frac{M^2}{2} R - \frac{1}{2} K^{\mu_1\mu_2\mu_3\mu_4}\nabla_{\mu_1} \tilde{A}_{\mu_2} \nabla_{\mu_3}\tilde{A}_{\mu_4},
\end{eqnarray}
where $R$ is the Ricci scalar, $K^{\mu_1\mu_2\mu_3\mu_4}$ is the most general covariant polynomial of $\tilde{A}$ consistent~\footnote{By consistent, we mean that there are no terms in which the number of $\tilde{A}$ fields can be reduced covariantly by directly imposing the constraint. For example, any term proportional to $\tilde{A}^\mu \nabla_\sigma \tilde{A}_\mu$ is not constraint consistent as can be seen by differentiating the constraint equation $\tilde{A}^2=v^2$.} with the Lagrange multiplier constraint,
\begin{equation}
 K^{\mu_1\mu_2\mu_3\mu_4} = \tilde{c_1} g^{\mu_1\mu_2}g^{\mu_3 \mu_4} + \tilde{c_2} g^{\mu_1\mu_3}g^{\mu_2 \mu_4}+ \tilde{c_3} g^{\mu_1\mu_4}g^{\mu_2 \mu_3}+ \frac{\tilde{c_4}}{v^2} \tilde{A}^{\mu_1}\tilde{A}^{\mu_2}g^{\mu_3 \mu_4},
\end{equation}
and $\tilde{c_i}$ with $i=1,\ldots 4$, are four free, dimensionless parameters. These are related to the usual $c_i$ parameters, as found in~\cite{update} for example, by 
\begin{equation}
\tilde{c_i} = (M/v)^2 c_i.
\end{equation}
We omit the term $R_{\mu\nu}\tilde{A}^\mu \tilde{A}^\nu$ because it is redundant; it is equal to a combination of the above terms under a field redefinition, up to a total derivative.  Taking $\tilde{c_i}\sim 1$, this model has two scales, the Lorentz violating aether VEV, $v$, and $M$ which is related to the reduced Planck mass by,
\begin{equation}
M^2 = M_{P}^2 + \frac{\tilde{c_1} + \tilde{c_4}}{2} v^2
\end{equation}
as can be found in the weak field, slow-motion limit of the classical equations of motion~\cite{Carroll:2004ai}. In this paper we will work in the parameter regime $v\lesssim M$. For a review of the constraints on the $\tilde{c_i}$ parameters see \cite{update} and references therein. For some recent constraints arising from considerations of stability, see \cite{carrollinst}.

Implicit in the use of the two-derivative Lagrangian (\ref{l2def}) in the solar-system or on larger scales is the assumption that the higher-derivative terms or quantum corrections are negligible. In this paper we make this statement precise using a dimensionally regulated quantum effective theory for Einstein-aether, making use of power-counting techniques. 

Using this formalism we are then able to investigate whether loop corrections can modify the scaling of the $\tilde{c_i}$ parameters. Recall that in QCD, for example, logarithmic corrections cause the value of the strong coupling to change from $\alpha_s \simeq 0.4$ at $1\, \text{GeV}$ to $\alpha_s \simeq 0.1$ at $100 \,\text{GeV}$, an effect which is subsequently measured~\cite{Amsler:2008zzb}. If there is a scale dependence of the $\tilde{c_i}$ in Einstein-aether theory, then it is of particular importance since the phenomenological couplings are compared with experiments which span many orders of magnitude in energy scale. For example, in the investigation of Cerenkov radiation~\cite{Elliott:2005va}, the most energetic cosmic ray particles considered have energies $10^{11}\, \text{GeV}$, which are many orders of magnitude greater than the typical curvature scales in the Solar-system. In turn, the length scales associated with the Solar-system are orders of magnitude smaller than those involved in cosmology. Nevertheless, observations on each of these scales have been used to constrain the two-derivative phenomenological parameters.

Perhaps one of the most remarkable features of Einstein-Aether theory is the possibility that $v\sim M$, even in the face of otherwise stringent constraints. This large value for $v$ is allowed because there are special combinations of the parameters, $c_i$, for which Einstein-aether theory is indistinguishable from general relativity in the PPN formalism~\cite{Foster:2005dk}. The conditions on these parameters are independent of the size of $v/M$, allowing $v\sim M$. In this regime one has a large gravitational modification. If there is a log-running of the two-derivative parameters then it would be surprising if these parameter choices were preserved under a change in energy scale. In this case a value of $(v/M)^2 \lesssim 10^{-7}$ would guarantee consistency with the PPN measurements without requiring delicately balanced parameters.

This paper is organised as follows.  We begin in section \ref{gr} with an introduction to quantum effective field theory techniques for gravitational theories, by reviewing the theory of quantised metric perturbations in pure gravity. We then apply this methodology to Einstein-aether theory in section \ref{ae} where we define the effective Lagrangian for Einstein-aether, and discuss its perturbative quantisation. In section \ref{power} we use power-counting techniques to estimate the size of a general Feynman diagram constructed in this theory, and in section \ref{running} we establish the hierarchy of a finite number of contributions to any given physical process. 

Having established that the effective theory possesses a controlled loop and derivative expansion we are then in a position to study the evolution of the parameters under the renormalisation group. We find that the low-energy, two-derivative parameters of this effective theory do not receive logarithmic corrections. This result follows in a straightforward manner from the property that all interaction vertices are derivative couplings.

Note that no steps in the above analysis could have been skipped; the non-running of the low-energy parameters follows from the structure of the perturbative low energy Lagrangian constructed in section \ref{ae}, which in turn can only be trusted given the power-counting result (sections \ref{power} and \ref{running}) which demonstrates that there is a controlled expansion.

In section \ref{directcouplings} we discuss the size of the direct aether-matter couplings, and the associated potential fine-tuning problem. Assuming a sufficiently small or vanishing two-derivative direct coupling term, as measured, we estimate the expected size of Lorentz violation which arises due to loop effects. This amounts to an estimation of the natural size for coupling constants of four-derivative dimension-six operators, for which there are also strong experimental constraints. This size depends on the ratio of scales $v/M$, hence it can be used to constrain the two-derivative $c_i$-parameters. We also comment on the use of a momentum-space cut-off regulator.

Related works on the subject include the semi-classical approaches \cite{Nacir:2007dw,LopezNacir:2008tx} which begin with a direct coupling of quantum matter to classical aether and metric, and then see how loops of matter renormalise the aether theory. Our interest is different here, we wish to study effects associated with quantised metric and aether perturbations. The phenomenology of tree-level diagrams in Einstein-aether theory has also been considered \cite{Elliott:2005va}.

%
%
%
%
%
%
%
%
%
%
%
%
%
%
%
%
%
%
\section{General relativity as an effective field theory}\label{gr}
In this section we provide a brief introduction to effective field theory by considering the quantum theory of metric perturbations in pure-gravity~\cite{Donoghue:1994dn,Donoghue:1995cz}. The Einstein-Hilbert Lagrangian contains a heavy mass scale, the (reduced) Planck mass, 
\begin{equation}\label{grlag}
\mathcal{L}_G= -\frac{M_{P}^2}{2} \sqrt{-g} R.
\end{equation}
Perturbing this Lagrangian about a weakly curved background solution $g_{\mu\nu} = \bar{g}_{\mu\nu} + M_{P}^{-1}\gamma_{\mu\nu}$, it is clear that canonically normalised metric perturbations carry inverse powers of the heavy mass. The graviton operator of lowest dimension is dimension four, the kinetic term, whilst all other operators are interacting and therefore carry increasing powers of $1/M_{P}$. These are irrelevant operators, signalling that the theory is non-renormalisable.

The presence of only non-renormalisable interactions means that divergences arising from loop integrals cannot be absorbed by redefining the couplings present in the Lagrangian. However, provided we are only interested in low energy scales, $E\ll M_P$, then effective field theory offers a consistent solution to this problem. Using a regularisation procedure which preserves the symmetries of the Lagrangian, loop divergences are guaranteed to have a counterterm if one adds all irrelevant local operators consistent with these symmetries.

Adding all irrelevant operators consistent with diffeomorphism invariance gives an effective action with an infinite number of coupling constants, which we have arranged in a derivative expansion\footnote{Note that we have not included the super-renormalisable cosmological constant term. This is determined by experiment to be incredibly small $\sim \sqrt{-g}\, 10^{-120} M_P^4$, and can be neglected if one is not interested in cosmological scales~\cite{Amsler:2008zzb}.}
\begin{equation}\label{eftgr}
S = \int{d^4x \sqrt{-g}\left(-\frac{M_{P}^2}{2} R + a_1 R_{\mu\nu} R^{\mu\nu} + a_2 R^2+ \ldots\right)}.
\end{equation}
The meaning of renormalisability in the effective field theory sense can then be seen most explicitly in the background field method, where explicit calculation confirms that one-loop~\cite{'tHooft:1974bx} and two-loop~\cite{Goroff:1985th} divergences are of the form of these higher derivative terms. Indeed, the effective theory for gravity is fully renormalisable in this sense~\cite{Gomis:1995jp}. 

Of course there are now an infinite number of terms and an infinite number of unknown coupling constants. Therefore if this theory is to be of any use one must find a consistent way to truncate the Lagrangian. To do so one must make a further assumption - that the coupling constants are \emph{natural}. Here this means that all coefficients are $\mathcal{O}(1)$ up to combinatorial factors and the scale $M_{P}$. Of course the size of the cosmological constant term is at odds with this assumption - a problem which is not resolved by the use of effective field theory. With this assumption, one can perform a straightforward power counting for the size of a general diagram $\mathcal{M}$ computed using a dimensional regulator and constructed with $\ell$ loops, $E_\gamma$ external graviton legs and $N^{(d)}$ vertices with $d$ derivatives,
\begin{equation}\label{grcount}
\mathcal{M} \sim M_P^{4-E_\gamma} \left(\frac{E}{M_P}\right)^{2+2\ell + \sum_{d=2}{N^{(d)}(d-2)}},
\end{equation}
where $E$ is the typical energy scale constructed invariantly from the external momentum in the graph, $E\ll M_{P}$. The hierarchy of contributions to this process may then be read off from (\ref{grcount}). The leading contribution to a particular process are tree level diagrams with two derivatives. Next-to-leading contributions are one-loop diagrams constructed with two derivative operators and tree level diagrams containing one four-derivative operator, and so on. The expansion (\ref{eftgr}) is therefore an expansion in $(E/M_{P})^2$, which breaks down if $E$ approaches $M_{P}$. Now we can see the utility of the effective approach; whilst we have an infinite number of undetermined coupling constants, this does not imply that the theory is not predictive. To compute a process of typical energy $E$, one simply has to specify the desired accuracy of the result, which determines where one can truncate the derivative expansion. This leaves a finite number of parameters, rendering the theory predictive. 

Note that one is only able to draw the above conclusions because there is one scale in the problem, $M_{P}$. This is a challenge presented by Einstein-aether, where the physics of the two-derivative Lagrangian depends on two scales.

In summary, the effective field theory of perturbative quantum gravity has the advantage of allowing quantum corrections to be calculated, provided the energy scales involved satisfy $E\ll M_{P}$.   At these energy scales, the theory is renormalisable as an effective field theory~\cite{Gomis:1995jp}.

%
%
%
%
%
%
%
%
%
%
%
%
%
%
%
%
%
%

\section{Einstein-aether effective theory}\label{ae}
In this section we construct an effective theory of quantised metric and aether perturbations for Einstein-aether. Specifically, we use a path-integral formulation with the Lagrange multiplier constraint directly imposed. This construction allows for the computation of quantum effects and their subsequent renormalisation.

We begin by writing down the most general action consistent with diffeomorphism invariance and the Lagrange multiplier constraint. It will be convenient to use a dimensionless aether field in what follows, $A\equiv \tilde{A} /v$, where $v$ is the size of the VEV for $\tilde{A}$. Throughout this paper any field with a tilde is canonically normalised. To represent the matter content of this theory we have added a scalar, $\phi$, for simplicity. We have,
\begin{eqnarray}
S&=&\int{d^4x \sqrt{-g}\left(\mathcal{L}[A^\mu,g_{\mu\nu},\phi]+\lambda\left(A^2-1\right)\right)},\label{barefull}\\
\mathcal{L}[A^\mu,g_{\mu\nu},\phi]&=&\sum_{d,a,s,i}{\alpha^{(i)}_{(d,a,s)}\mathcal{O}_{(i)}^{(d,a,s)}[A^\mu,g_{\mu\nu},\phi]},\label{barelag}
\end{eqnarray}
where $\mathcal{O}_{(i)}^{(d,a,s)}$ denotes a constraint consistent, diffeomorphism invariant term which can be constructed from $d$ derivatives, $a$ aether fields, $s$ matter fields and the metric. There are potentially a number of different terms with the same counting due to differing Lorentz structures, and so we add a further label, $i$, which enumerates all possible terms. Note that we have not excluded the potentially problematic direct $\phi$, $A^\mu$ couplings, for example, $A^\mu \partial_\mu \phi A^\nu \partial_\nu \phi$; we have included all terms. 

As in the pure-gravity example, there are an infinite number of terms in the Lagrangian (\ref{barelag}). We therefore need a way to truncate this theory, and this will be the focus of sections \ref{power} and \ref{running}. In this section we will define the perturbations and the quantisation procedure using a dimensional regulator. The main reason that we use a dimensional regulator is to preserve the diffeomorphism symmetry. This regulator is used in pure-gravity~\cite{'tHooft:1974bx,Goroff:1985th}, and in theories which are structurally similar to Einstein-aether in the absence of gravity, for example in chiral perturbation theory~\cite{Gasser:1983yg}. Dimensional regularisation allows for a controlled truncation of the effective theory as we shall show, whilst a momentum-space cut-off gives an uncontrolled effective expansion. We discuss this issue in section \ref{directcouplings}.

The coupling constants have mass dimension,
\begin{equation}
\left[\alpha^{(i)}_{(d,a,s)}\right] = 4-d-s,
\end{equation}
and with the assumption of naturalness are to be built from the scales in this theory, together with some combinatorial factors and an $\mathcal{O}(1)$ number. The scales that appear in these couplings are dependent on the renormalisation point $\mu$ at which the theory is defined. Here for convenience we will focus on the characteristic energy scale associated with the curvatures in the solar-system, or  on larger length scales. Hence it is most convenient to consider the effective theory defined at a renormalisation point well below the mass of the lightest particle in the standard model, which we denote by $m$. We take,
\begin{equation}
E \,\sim \,\mu\, \ll \, m \,\lesssim\, v\, \lesssim \,M.
\end{equation}
That is, we have `integrated out' all massive particles through a matching procedure~\footnote{For a discussion of \emph{continuum effective field theory} see for example~\cite{Georgi:1994qn,Burgess:2007pt}} on to the effective Lagrangian (\ref{barelag}). The results obtained within this regime are easily extended to higher energy scales, $m \lesssim E \ll v \lesssim M$, and we discuss this in section \ref{running}.

At low energies only virtual processes of the massive particles are important and this is exactly what the higher-derivative operators in the effective Lagrangian (\ref{barelag}) are chosen to achieve through the procedure of matching. Hence they will typically have inverse powers of the scale $m$ in their coefficients. Also important at low energies are of course the massless degrees of freedom. The two-derivative Lagrangian (\ref{l2def}) explicitly defines the massless gravitational and aether modes. Additionally we have the massless remains of the standard model. For simplicity we will idealise this matter content by using the scalar field above, $\phi$, taking it to be massless and to not self-interact at two-derivative order.

We perform a $3+1$ split of $A$, and a decomposition of the metric about flat space,
\begin{eqnarray}
A &=& \sigma \partial_0 +  \pi^i \partial_i,\label{adecomp} \\
g_{\mu\nu} dx^\mu dx^\nu &=& (\eta_{\mu\nu} +  \gamma_{\mu\nu})dx^\mu dx^\nu,
\end{eqnarray}
Treating the $\pi$ and $\gamma$ fields perturbatively picks out one effective field theory in particular, corresponding to choosing a coordinate frame in which the background value of the aether field is purely time aligned.  In principle one may  choose to treat a different set of fields as perturbations, corresponding to perturbation theory about a different background Lorentz frame, and a different effective field theory~\footnote{It has been recently pointed out \cite{carrollinst} that if the observer frame is highly boosted with respect to the aether rest frame, higher derivative terms can be lower order in the energy expansion than naively expected. Again, we stress that this is a different effective field theory altogether, and one would have to perform a new power counting analysis.
}.  However the frame we have chosen here is the one in which the phenomenology of Einstein-aether has been investigated and compared to experiment, for example in the PPN formalism~\cite{Eling:2003rd,Foster:2005dk}.

The approach we take here is to eliminate $\sigma$ from the Lagrangian using the Lagrange multiplier constraint, which may be performed formally in the path-integral by integrating over $\lambda$, 
\begin{equation}\label{sdelta}
Z=\int \mathcal{D}A\,\mathcal{D}\gamma\,\mathcal{D}\lambda\; e^{i\int d^4x \mathcal{L}+ \lambda \left(A^2 -1\right)}\propto \int \mathcal{D}A\,\mathcal{D}\gamma\; \delta\left(A^2 -1\right)e^{i\int d^4x \mathcal{L}}.
\end{equation}
This path-integral requires gauge fixing, and there is the same amount of diffeomorphism degeneracy here as in the case of general relativity. One may therefore use a pure-gravitational gauge breaking term of choice, such as the Prentki gauge~\footnote{See for example the lectures by Veltman in~\cite{Balian:1976vq}}. A purely gravitational gauge fixing of course leads to the introduction of the usual Faddeev-Popov ghost terms, and consequently, graviton-ghost interaction vertices. Being quadratic in the ghost field, these ghost terms have the same power counting properties as pure graviton vertices. Hence, as is the case for power counting in pure gravity (see for example~\cite{Donoghue:1995cz}), we do not need to include them explicitly in our analysis. One should therefore bear in mind that in this paper we have omitted Feynman diagrams which contain internal ghost lines and these would have to be included if one went beyond power counting. 

Continuing from (\ref{sdelta}) we may integrate over one of the fields in the delta-functional. In particular we choose to integrate over $A^0\equiv\sigma$. Employing the identity
\begin{equation}
\int \mathcal{D}\sigma \, Det\left(\frac{\partial{A^2}}{\partial\sigma}\right)\delta\left(A^2-1\right)=1,
\end{equation}
and the expression for $\sigma$ given by solving the constraint equation $A^2=1$,
\begin{equation}
\sigma=g^{-1}_{00} \left(-g_{0i} \pi^i \pm \left[g_{00} + (g_{0i}\pi^i)^2 - g_{00}g_{ij} \pi^i \pi^j\right]^{\frac{1}{2}}\right),
\end{equation}
we obtain,
\begin{equation}\label{zdet}
Z\propto \int \mathcal{D}\pi\mathcal{D}\gamma \left[Det\left(\pm 2(g_{00}  + (g_{0i}\pi^i)^2 - g_{00} g_{ij} \pi^i \pi^j)^{\frac{1}{2}}\right)\right]^{-1}e^{i\int d^4x \mathcal{L}[\pi^i,\gamma_{jk},\phi]},
\end{equation}
where the plus and minus roots correspond to a choice of future or past-directed $A$, respectively. There will of course be Goldstones associated with the vacuum Lorentz violation, and these correspond to a mixture of the new $\pi$ fields and the metric perturbations. There are no massive modes, since the constraint forces the aether to lie exactly on the unit hyperboloid. 

The determinant may be absorbed into the Lagrangian by the use of the functional identity $Det\, M = \exp{Tr \log{M}}$. This will result in new, quartically divergent operators in the $\pi$ fields. An analogous determinant appears in the perturbative quantisation of the non-linear sigma model, see for example~\cite{Gerstein:1971fm,Weinberg:1995mt}. In this case the determinant is known to account for the non-linearly realised symmetry, retaining the correct invariance of the functional integration measure. In turn, it gives rise to new power-law divergent operators in the Lagrangian which cancel against loop integrals which do not respect the symmetry of the Lagrangian. Here we work with dimensional regularisation for which there are no power-law divergences, and the determinant will not contribute new interaction terms. If a cutoff regulator were to be used instead, we expect that the determinant here plays an analogous role to the determinant found in the non-linear sigma model, and such terms must therefore be included.

All operators in this theory are the result of expanding covariant, constraint consistent terms. As an example, consider a kinetic term for the aether field in the 2-derivative Lagrangian. Ignoring the index structure and order-one numbers, its perturbative expansion will be of the form,
\begin{equation}\label{pertform}
\sqrt{-g} \,\tilde{c} v^2\, \nabla A \nabla A \sim \tilde{c} \left( \partial\tilde{\pi} \partial \tilde{\pi} + \left(\frac{v}{M}\right)\partial\tilde{\gamma} \partial\tilde{\pi} + \left(\frac{v}{M}\right)^2 \partial \tilde{\gamma} \partial \tilde{\gamma} + \frac{1}{v} \tilde{\pi} \partial \tilde{\pi} \partial \tilde{\pi} + \frac{1}{M} \tilde{\pi}\partial\tilde{\pi} \partial \tilde{\gamma} + \ldots\right),
\end{equation}
where the perturbations have been canonically normalised, 
\begin{equation}
\tilde{\pi} \equiv v \pi,  \qquad \tilde{\gamma}\equiv M\gamma.
\end{equation}
Thus once the integration over $\lambda$ and $\sigma$ has been performed, one covariant term in the Lagrangian gives rise to a finite number of marginal operators and an infinite tower of irrelevant operators of increasing mass dimension, with \emph{one} coupling constant up to factors of $v/M$. Then, if our regularisation and renormalisation procedure preserves the symmetries of the covariant form of the Lagrangian, then all logarithmic divergences may be absorbed by these couplings. For example, the effective theory of pure gravity discussed in section \ref{gr} has been shown to be renormalisable in this sense~\cite{Gomis:1995jp}. We do not have an analogous proof for Einstein-aether, and we assume this property.

Note that in general the propagator for the $\gamma$ and $\pi$ perturbations is not block-diagonalised, that is,
\begin{equation}
 \left<\tilde{\gamma}_{\mu\nu} \tilde{\pi}^i\right>\neq 0. 
\end{equation}
This of course depends on the specific numerical relationships between the coefficients of the marginal operators, and the chosen gauge fixing condition. For the purposes of power counting it will be convenient to retain the perturbations as defined here, because we are interested in counting the mass scales $v$ and $M$.

%
%
%
%
%
%
%
%
%
%
%
%
%
%
%
%
%
%
\section{Power Counting}\label{power}
As discussed in section \ref{gr}, an effective theory with an infinite number of coupling constants -- such as that outlined in the last section -- must be supported by a power-counting argument which allows for a controlled truncation of the Lagrangian based on the accuracy of the result required. This is necessary for the theory to be predictive. In this section we take the first step; we perform power-counting to estimate the size of any Feynman diagram. This is supplemented by section \ref{running}, where we show that this leads to a truncatable Lagrangian.

We begin by considering a general Feynman diagram in momentum space with $\ell$ loops and $E_\pi, E_\gamma$ and $E_\phi$ external aether, graviton and matter lines respectively. We denote its amplitude by $\mathcal{M}$, which has the mass dimension
\begin{equation}\label{mdim}
\left[ \mathcal{M} \right]=4-E_\pi -E_\gamma - E_\phi.
\end{equation}
As usual we will not include factors of the momentum arising from external leg propagators or the momentum-conservation delta function in $\mathcal{M}$. Of course, if one wanted the size of the $(E_\phi+E_\pi+E_\gamma)$-point function itself, then these would have to be included in addition to the amplitude $\mathcal{M}$. Since we are working in dimensional regularisation any renormalisation scale $\mu$ will not contribute to the dimension of $\mathcal{M}$. The dimension (\ref{mdim}) is therefore accounted for entirely by typical light external momenta $E$, and the heavy mass scales $v, M$ and $m$.

Calculating $\mathcal{M}$ by combining contributions from vertices and internal lines generically results in a complicated multi-integral expression. However an estimate for its size is easily deduced since the leading powers of $v, M, m$ necessarily factor out. Our first step will be to count these heavy mass factors, which are all contributed by couplings at vertices and by the non-canonical internal propagators. 

As previously stressed, the sizes of the higher derivative couplings, and some of the two-derivative couplings are not known. Ultimately these are to be determined by experiment. We first discuss the two-derivative couplings, for which the theory is currently over-parametrised, since we may always perform a simultaneous rescaling of the aether field and redefinition of $v$ to absorb one of the couplings. There are four possible terms which can yield a kinetic term for the spatial aether perturbation, $\pi$. These are,
\begin{equation}
\begin{split}
\alpha^{(1)}_{(2,2,0)} \,\sqrt{-g} \nabla_\mu A^\mu \nabla_\nu A^\nu, \qquad\quad  & \alpha^{(2)}_{(2,2,0)} \,\sqrt{-g}\nabla_\mu A^\nu \nabla_\nu A^\mu, \\
\alpha^{(3)}_{(2,2,0)} \,\sqrt{-g} \nabla_\mu A^\nu \nabla^\mu A_\nu, \qquad\quad  & \alpha_{(2,4,0)} \,\sqrt{-g} A^\mu A^\nu \nabla_\mu A_\rho \nabla_\nu A^\rho.
\end{split}
\end{equation}
We use the $\pi$ kinetic term to \emph{define} the scale $v$, that is, we take 
\begin{equation}
\alpha^{(i)}_{(2,2,0)}\sim \alpha_{(2,4,0)} \sim v^2. 
\end{equation}
Additionally the Ricci scalar term defines the scale $M$, and we canonically normalise $\phi$. 

Hence for the two-derivative Lagrangian the only terms not yet discussed are direct couplings between the aether and matter. The matter Lagrangian here is massless and does not self-interact and so is invariant under a constant shift, $\phi\to \phi+c$. Demanding that this shift symmetry is not broken by direct aether matter couplings then restricts the derivatives to act on the $\phi$ fields and not on the aether fields in these terms. Hence the only two-derivative direct aether-matter coupling with these symmetries contains exactly two aether fields. This term contains a Lorentz-violating contribution to the $\phi$ propagator at leading order in its perturbative expansion,
\begin{equation}\label{LV}
\alpha_{(2,2,2)} \sqrt{-g} A^\mu A^\nu \partial_\mu \phi \partial_\nu \phi  = \alpha_{(2,2,2)} \partial_0 \phi \partial_0 \phi + \ldots
\end{equation}
where the ellipses denote higher dimension operators. Given the experimental bounds on the smallness of direct aether-matter coupling constants (see for example~\cite{Coleman:1998ti,Colladay:1996iz,Colladay:1998fq, Mattingly:2008pw, Mattingly:2005re}), we impose the minimal phenomenological requirement that this marginal term is not larger than the contribution from the Lorentz-invariant piece of the $\phi$ propagator. That is, we require
\begin{equation}\label{minimal}
\alpha_{(2,2,2)} \leq 1.
\end{equation}
This concludes a discussion of the size of the two-derivative Lagrangian couplings. 
 
Moving on to the higher-derivative couplings, we include all operators which are consistent with the required symmetries, using the dimensionless fields $g_{\mu\nu}$, $A$ and the unit mass-dimension $\phi$. Our goal here is to establish whether or not the theory has a controlled effective expansion, and to achieve this we construct these couplings with inverse powers of the mass scale $m$,
\begin{eqnarray}\label{max}
\alpha_{(d,a,s)} & \sim & \frac{1}{m^{(d-4) +s}}, \qquad d\geq 4, \qquad m\lesssim v \lesssim M.
\end{eqnarray}
We choose these values because they constitute an upper limit on their natural sizes, since any loop corrections they receive will be smaller by factors of $m/v$ or $m/M$. If these couplings are indeed smaller than (\ref{max}), as may be found through experiment, this will only improve the effective expansion.

We now look at how mass scales enter in the internal propagators. In momentum space, the aether and metric marginal operators in the Lagrangian take the following form,
\begin{equation}
\left(
\begin{array}{c}
\vec{\gamma}(p)\\ 
\vec{\pi}(p)
\end{array}
\right)^\mathrm{T}
\left(
\begin{array}{cc}
M^2 \mathbf{A}^{\mu\nu}& v^2 \mathbf{B}^{\mu\nu}\\ 
v^2 \mathbf{B}^{\mu\nu, \mathrm{T}} & v^2 \mathbf{C}^{\mu\nu}
\end{array}
\right)
\left(
\begin{array}{c}
\vec{\gamma}(-p)\\ 
\vec{\pi}(-p)
\end{array}
\right)p_\mu p_\nu,
\end{equation}
where $\vec{\gamma}$ is a vector of the ten components of the metric perturbations, $\vec{\pi}$ is a vector of the three spatial aether components and $\mathrm{T}$ denotes the transpose on these indices. The matrices $\mathbf{A}^{\mu\nu},\mathbf{B}^{\mu\nu}$ and $\mathbf{C}^{\mu\nu}$ account for order-one numbers and the structure of the marginal operators in terms of the ten components of $\vec{\gamma}$ and the three components of $\vec{\pi}$. The spacetime indices of these matrices contain the structure of the momentum dependence of these operators, and it is worth emphasising that they are not Lorentz tensors. To determine the typical sizes for the inverse of this matrix we take the limit where $v/M$ is small. Then,
\begin{equation}
\begin{split}
&\left(
\begin{array}{cc}
M^2 \mathbf{A}^{\mu\nu}& v^2 \mathbf{B}^{\mu\nu}\\ 
v^2 \mathbf{B}^{\mu\nu, \mathrm{T}} & v^2 \mathbf{C}^{\mu\nu}
\end{array}
\right)^{-1} = \left[
U
\left(
\begin{array}{cc}
\mathbf{A}^{\mu\nu}& \frac{v}{M} \mathbf{B}^{\mu\nu}\\ 
\frac{v}{M}\mathbf{B}^{\mu\nu, \mathrm{T}} & \mathbf{C}^{\mu\nu}
\end{array}
\right)
U\right]^{-1}\\
&\qquad=U^{-1}
\left(
\begin{array}{cc}
\mathbf{A}^{-1}_{\mu\nu} + \mathcal{O}\left(\frac{v}{M}\right)^2 & -\frac{v}{M}\left( \mathbf{A}^{-1}\mathbf{B} \mathbf{C}^{-1}\right)_{\mu\nu}+ \mathcal{O}\left(\frac{v}{M}\right)^2\\ 
-\frac{v}{M}\left( \mathbf{C}^{-1}\mathbf{B}^{ \mathrm{T}} \mathbf{A}^{-1}\right)_{\mu\nu} + \mathcal{O}\left(\frac{v}{M}\right)^2 & \mathbf{C}^{-1}_{\mu\nu}+ \mathcal{O}\left(\frac{v}{M}\right)^2
\end{array}
\right)
U^{-1}\\
&\qquad=\quad\;\left(
\begin{array}{cc}
\frac{1}{M^2}\mathbf{A}^{-1}_{\mu\nu}+ \mathcal{O}\left(\frac{v^2}{M^4}\right) & -\frac{1}{M^2}\left( \mathbf{A}^{-1}\mathbf{B} \mathbf{C}^{-1}\right)_{\mu\nu}+ \mathcal{O}\left(\frac{v}{M^3}\right)\\
-\frac{1}{M^2}\left( \mathbf{C}^{-1}\mathbf{B}^{ \mathrm{T}} \mathbf{A}^{-1}\right)_{\mu\nu}+ \mathcal{O}\left(\frac{v}{M^3}\right)& \frac{1}{v^2}\mathbf{C}^{-1}_{\mu\nu}+ \mathcal{O}\left(\frac{1}{M^2}\right)
\end{array}
\right),
\end{split}
\end{equation}
where,
\begin{equation}
U\equiv
\left(
\begin{array}{cc}
M \mathbbm{1} & 0\\ 
0 & v\mathbbm{1}
\end{array}
\right).
\end{equation}
In general, the off-diagonal blocks of this matrix, $\left( \mathbf{A}^{-1}\mathbf{B} \mathbf{C}^{-1}\right)_{\mu\nu}$ and $\left( \mathbf{C}^{-1}\mathbf{B}^{ \mathrm{T}} \mathbf{A}^{-1}\right)_{\mu\nu}$, will not vanish. Then, the heavy mass scales entering into the propagator sizes for the non-canonically normalised perturbations $\pi$ and $\gamma$ will be given by,
\begin{equation}
\begin{split}\label{props}
\gamma\; \includegraphics{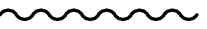}\; \gamma & \qquad\sim\qquad\frac{1}{M^2 p^2} \left(1+\mathcal{O}\left(\frac{v}{M}\right)^2\right)\\
\pi\; \includegraphics{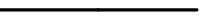} \;\pi & \qquad\sim\qquad\frac{1}{v^2 p^2}\left(1+\mathcal{O}\left(\frac{v}{M}\right)^2\right)\\
\pi\; \includegraphics{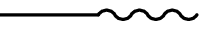}\; \gamma & \qquad\sim\qquad\frac{1}{M^2 p^2}\left(1+\mathcal{O}\left(\frac{v}{M}\right)\right)
\end{split}
\end{equation}
In the special case where $v\sim M$, there will be only one scale, and hence the above propagators also give the correct counting in this regime, despite their derivation in the regime where $v \ll M$. Hence we assume that they provide the correct counting for $v \lesssim M$. 

With a different choice of gauge for the perturbations it is possible that the off-diagonal blocks could be made to vanish. However, any physical results will be independent of the gauge fixing condition, and so we are free to work in the generic case for which the above counting holds. Here the $p^2$ factor represents the contribution of the typical light scale of the momentum in the propagator, and in general this will not be a Lorentz invariant quantity. Note however that the momentum dependence is still quadratic~\footnote{This is to be contrasted with other theories which manifestly break Lorentz invariance such as the ghost condensate model~\cite{ArkaniHamed:2003uy} where dispersion relations become quartic.} in both the time, and the spatial components of $p$.

With the mass scales contributed by the propagators determined in (\ref{props}), one could explicitly count the mass-scale contribution from the $\left<\pi\gamma\right>$ propagator simply by counting how many times it appears in the graph. One must be careful however, since when $\pi$ and $\gamma$ are both canonically normalised, the propagator $\left<\tilde{\pi}\tilde{\gamma}\right>$ still carries a factor of $v/M$. Ordinarily, specifying the external leg content determines completely which propagator is used in the external leg. However, here we must account for the possibility that the `species' can change in the external leg, which subsequently contributes further factors of the ratio $v/M$. A useful mnemonic to keep track of the factors of $v$ and $M$ introduced by the $\left<\pi\gamma\right>$ propagator is to count it as a `vertex' which has one $\gamma$ leg and one $\pi$ leg, and which carries a the coupling constant $v^2 p^2$. We emphasise that this is not a Feynman vertex, but a combinatorial tool. Hence it is subject to the additional rule that it may be included only once in any given line of the Feynman diagram. We then have,
\begin{equation}
\pi\;\includegraphics{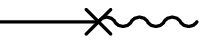}\;\gamma \quad \sim\quad \frac{1}{v^2 p^2} \cdot v^2 p^2 \cdot \frac{1}{M^2 p^2} \quad\sim\quad \pi\; \includegraphics{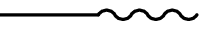}\; \gamma
\end{equation}
as required. 

We enumerate the total number of vertices by $N$, which we then partition according to the number of derivatives, $d$, the number of matter fields, $s$, and the number of factors of the aether field, $a$, present in the original covariant term from which the operator is derived,
\begin{equation}
N=\sum_{d,a,s,}{N^{(d,a,s)}} + \mathcal{V},
\end{equation}
where $\mathcal{V}$ counts the number of species-changing `vertices', as described above. We denote the number of internal propagators in our diagram by $I_{\gamma},  I_{\pi}$ and $I_{\phi}$.

In the general case, summing up contributions from heavy mass scales we obtain,
\begin{equation}\label{messycount}
\begin{split}
\mathcal{M} \sim  \qquad & M^{2N^{(2,0,0)} - 2I_\gamma} (\alpha_{(2,2,2)})^{N^{(2,2,2)}} v^{2 N^{(2,2,0)} + 2 N^{(2,4,0)} + 2 \mathcal{V} - 2I_\pi} \\
& \times \;  m^{\sum_{d\geq 4, a,s} N^{(d,a,s)} \left(4-d-s\right)} E^D v^{-E_\pi} M^{-E_\gamma}.
\end{split}
\end{equation}
The remaining mass dimension of $\mathcal{M}$ is accounted for by $D$ powers of the light scale $E$, and the last two terms account for the non-canonical normalisation of $\gamma$ and $\pi$. We may now determine $D$ by the requirement that $\mathcal{M}$ have the correct mass dimension (\ref{mdim}). After some manipulation, and use of the topological identity, $I_{\gamma}+I_{\pi}+I_{\phi} = \ell + N -1$ we find,
\begin{equation}\label{D}
D = 2 + 2\ell + \sum_{d,a,s}{N^{(d,a,s)} (d-2)}.
\end{equation}
There are now no unknown quantities in (\ref{messycount}), and we have therefore estimated the size of any Feynman diagram in this effective theory.

The equation (\ref{D}) is the usual effective field theory result found by Weinberg in the context of Chiral Lagrangians~\cite{Weinberg:1978kz}. It is the same as the $E$-counting result in the pure gravity case (\ref{grcount}). Increasing the number of loops in the diagram or the number of derivatives at vertices necessarily increases the power of $E$, which, in a theory with only one mass scale has the simple interpretation of increasing suppression by factors of $E/M$. However, such a result is meaningless in the presence of extra scales if one cannot show that there is a controlled expansion in ratios of those scales too, for example, $v/M$. We will address this issue in the next section.

As an example of the counting performed in this section, consider the following diagram which we have constructed using the interaction vertices:  $\sqrt{-g}  \nabla \tilde{A} \nabla \tilde{A} \supset v^2 \pi^2 \partial \pi \partial \pi,\,$ $ v^2 \gamma\gamma \partial \gamma \partial \pi$, 
\begin{equation}
\includegraphics{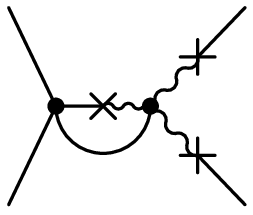}
\end{equation}
This graph has five vertices in total, $N=5$, which are made up from two interaction vertices from the two-derivative Lagrangian $N^{(2,2,0)}=2$ (filled circles), and three species changing `vertices', $\mathcal{V}=3$ (crosses). The internal line content is $I_{\phi}=0,\,I_{\pi}=2,\,I_\gamma=3$ and the external line content is $E_\phi=E_\gamma=0,\,E_\pi=4$. Inserting these values into (\ref{messycount}) we estimate the size of this diagram to be,
\begin{equation}
\mathcal{M} \sim  \left(\frac{E}{M}\right)^4 \left( \frac{v}{M}\right)^2.
\end{equation}

\section{Validity of the effective expansion and logarithmic corrections}\label{running}
Using the power-counting results of the last section we show that the effective theory defined in section \ref{ae} can be truncated to a Lagrangian with a finite number of terms, with the number of terms dependent on the order of accuracy required. As discussed, this is necessary for the theory to be predictive. Once this is established we are then able to estimate the scale-dependence of the couplings  arising due to logarithmic corrections.

In order to determine whether there is a controlled expansion we consider the size of a general diagram (\ref{messycount}). One can perform a straightforward re-arrangement of the summation (\ref{messycount}) to obtain,
\begin{equation}\label{type1}
\begin{split}
\mathcal{M} \sim m^{4-E_\pi-E_\gamma-E_\phi} \left[\frac{E}{m}\right]^D & \left[\frac{m}{v}\right]^{2I_\pi + E_\pi+ 2I_\gamma + E_\gamma- 2N^{(2,2,0)} - 2N^{(2,4,0)} -2N^{(2,0,0)}-2\mathcal{V}} \\
\times & \left[\frac{v}{M}\right]^{2I_\gamma + E_\gamma -2N^{(2,0,0)}} \left[\alpha_{(2,2,2)}\right]^{N^{(2,2,2)}}
\end{split}
\end{equation}
All quantities in square brackets in this expression are $\mathcal{O}(1)$ or smaller, and hence there is a well defined expansion if the power to which they are raised is bounded from below. If they are not bounded from below then they can provide large ratios of scales which could offset the $E/m$ suppression arising from increasing loop or derivative order. In this case diagrams from arbitrarily high derivative order can contribute and the derivative expansion cannot be truncated.

It is easily seen that the power of $\alpha_{(2,2,2)}$ is bounded from below. For powers of $v/M$ we note that the total number of graviton `ends' in the graph ($2I_\gamma + E_\gamma$) cannot be less than the total graviton valency contributed by the Ricci-scalar term, which is greater than two per vertex, that is,
\begin{equation}
2I_{\gamma} + E_\gamma > 2 N^{(2,0,0)}.
\end{equation}
The remaining power of $m/v$ is also bounded from below, as can be seen by a similar argument; each vertex in the graph arising from the expansion of $\sqrt{-g} \nabla A \nabla A$ , $\sqrt{-g} AA \nabla A \nabla A$ or $\sqrt{-g} R$, must have a combined total graviton and $\pi$ valency of at least two. Therefore the total graviton and $\pi$ valency in the graph cannot be less than this amount, that is,
\begin{equation} 
2I_\pi + E_\pi+ 2I_\gamma + E_\gamma \geq 2\left( N^{(2,2,0)} + N^{(2,4,0)} + N^{(2,0,0)}+\mathcal{V}\right).
\end{equation}

We have now established the existence of a controlled expansion. That is, given the energy scale of interest and the order of accuracy required, there are a finite number Feynman diagrams to evaluate. This analysis justifies the use of the two-derivative Lagrangian in describing processes at sufficiently low energies. We remind the reader that in order to obtain the power-counting result (\ref{type1}), we constructed the higher derivative couplings using inverse powers of the scale $m$. The coupling constants as determined experimentally could be smaller, in which case only vertex contributions are affected and the effective expansion would be improved.

We are now in a position to discuss the full implications of equation (\ref{D}), in the context of the renormalisation group. A consequence of (\ref{D}) is that a one-loop graph necessarily scales as at least $E^4$. For example, consider the matrix element for  $\tilde{\pi}-\tilde{\pi}$ scattering up to order $E^4$, which we denote by 
\begin{equation}
M(E) \equiv \sum_i\mathcal{M}_i,
\end{equation} 
where $i$ runs over all  graphs which can contribute. We work in a subtraction scheme, for example $\overline{MS}$. Once any divergences from loop integrals have been absorbed into the couplings, suppressing $\mathcal{O}(1)$ numbers and any Lorentz structure, we find
\begin{equation}\label{matrixel}
\begin{split}
M(E) \sim \quad & \frac{E^2}{v^2} \left[\sum_{j=1}^2 b^{(2)}_j \,\alpha_{(2,2 j,0)}(\mu) \right]\\ 
+ & E^4 \left[ \mathcal{A}\,\sum_{j=1}^4 b^{(4)}_j \alpha_{(4,2 j,0)}(\mu) + \frac{1}{v^4}\log{\left(\frac{E}{\mu}\right)} \right]+ \mathcal{O}\left(E^6\right),
\end{split}
\end{equation}
where $b^{(2)}_j$ and $b^{(4)}_j$ are dimensionless $\mathcal{O}(1)$ constants, and the sum runs over the number of aether fields in the covariant terms which can contribute to this process. The maximum number of aether-fields which can occur in a covariant term containing $d$-derivatives is $2d$, as any more cannot be constraint consistent. $\mathcal{A}$ is a constant of mass dimension $[\mathcal{A}]=-4$, which depends on details of the size of the higher-derivative couplings. For the assignment of higher-derivative couplings used above, $\mathcal{A} \sim v^{-2} m^{-2}$, but in practise it can take a smaller value and the theory will still have a good expansion. The expression for $M(E)$ must be independent of the renormalisation point, $\mu$, 
\begin{equation}\label{condition}
\frac{d}{d\mu} M(E) = 0.
\end{equation}
Hence in general the four derivative couplings evolve under the renormalisation group. The two derivative couplings do not receive radiative corrections \footnote{
The order $E^2$ piece of equation (\ref{matrixel}) together with the condition (\ref{condition}), suggests that it is possible for the two-derivative coupling constants to evolve on some hyper-surface in parameter space. This would mean that the two-derivative Lagrangian was over parametrised. However, the Lorentz structures associated with the different contributions $\alpha_{(2,2,0)}$ and $\alpha_{(2,4,0)}$ are in general different, and this is not captured by (\ref{matrixel}). Furthermore, if this invariance of the Lagrangian exists then it would be seen classically, and no such symmetry of the two-derivative Lagrangian has been found.}, and hence we conclude that the two-derivative parameters do not have any scale dependence. This justifies the application of experimental constraints to these parameters, despite their spanning of many orders of magnitude in energy scale.

We now comment on the regime in which the above results were obtained. One can work in the effective theory above the mass scale $m$, for which there is at least one massive particle which can contribute to the running of the two-derivative gravitational parameters. It is easily seen that the only possible one-loop corrections to the two-derivative gravitational/aether parameters involve at least one mass term and hence are at least $(m/M)^2$ suppressed, instead of $(E/M)^2$ as in the massless case. These corrections are therefore negligible for a standard-model matter content.

Of course there are very heavy masses in the solar-system, the Sun being one example. The quantum effective theory for non-relativistic masses in the case of General Relativity has been studied in \cite{Donoghue:1996mt}. Despite appearances, there is still a good effective expansion for this theory, which is more clearly seen when one adopts a non-covariant, physical gauge. The expansion parameter was found to be the same as in the vacuum case, that is, $(E/M_p)^2$. It is unclear whether Einstein-aether theory possesses a controlled expansion in the presence of non-relativistic masses and we leave this as an open question.

\section{Direct aether-matter couplings}\label{directcouplings}
In this section we use the effective theory to estimate the expected size of Lorentz violation due to loop effects. This is intimately related to the size of the coefficients of direct-aether matter couplings, and we outline a way one can use these estimates to bound the two-derivative $c_i$ parameters.

The two-derivative direct aether-matter couplings are commonly set to zero (for the matter content above, this coupling is $\alpha_{(2,2,2)}$). This is due to strict constraints coming from the observed accuracy of Lorentz invariance in the matter sector, together with the fact that this coupling modifies the matter dispersion relation in a Lorentz violating way (see \ref{LV}). 

Here, to make progress, we assume that there is some property of the underlying theory -- a symmetry perhaps --  which predicts a sufficiently small or zero direct coupling at two-derivatives in this low energy theory. Supersymmetry provides one such candidate ~\cite{GrootNibbelink:2004za,Bolokhov:2005cj}. In section \ref{running} we saw that the parameters of the two-derivative Lagrangian do not receive radiative corrections when below the scale $m$ and only small corrections above, and so the approximate size of this coupling does not change with the renormalisation point in this theory. Hence by setting this coupling to zero at one renormalisation point we will find it remains zero at larger length scales. Of course there can still be Lorentz-violating effects in matter processes which arise through loop effects, and higher derivative terms. We also know that the higher derivative terms can be scale dependent and so in general it is inconsistent to set these to zero.

Hence we allow for higher derivative direct couplings, for which there are significant experimental constraints in the four-derivative case, though these are less severe than for the two-derivative operators. For example, recent work for fermionic operators~\cite{Maccione:2009ju} constrain these kinds of dimension-six couplings to be, $\alpha \lesssim 10^{-6} M_p^{-2}$. For simplicity we retain the use of scalar matter, $\phi$. At four derivative order we have two kinds of direct-coupling term which, once expanded, give rise to a dimension-six operator for $\phi$ alone. These are, 
\begin{equation}\label{dim6}
 \alpha_{(4,2,2)} \sqrt{-g} A A \nabla\partial \phi \nabla \partial \phi, \qquad \alpha_{(4,4,2)} \sqrt{-g} A A A A \nabla\partial \phi \nabla \partial \phi,
\end{equation}
where we have suppressed Lorentz indices. 

To estimate the size of these couplings we use naive dimensional analysis (NDA). This method was first proposed in~\cite{Manohar:1983md} for the chiral theory of low energy QCD, which provides a way to estimate geometrical factors of $4\pi$ in the non-renormalisable couplings. In this section we use the NDA approach to estimate factors of the scales $v$ and $M$.

The basic idea behind NDA is that the coupling constants are set to be the same size as the largest radiative correction they each receive. The NDA result then gives the typical size of the coupling which one would generate under a change in the renormalisation point, $\mu$.

To study radiative corrections to $\alpha_{(4,2,2)}$ and $\alpha_{(4,4,2)}$ we are free to study any process to which they contribute at tree-level. Since we are using power-counting to determine the size of these diagrams, any cancellations which occur between individual loop graphs cannot be seen. Power counting therefore gives the best estimate for the corrections to these couplings when we study the process for which the leading loop correction is the smallest. If larger corrections are found by studying other processes, covariance of the unperturbed Lagrangian then guarantees that there will be appropriate cancellations.

The processes which give the best estimate for the size of loop corrections to (\ref{dim6}) are those which involve at least one external $\tilde{\pi}$ leg. Here we consider the process $E_\pi = E_\phi =2$, $E_\gamma = 0$, for which we can construct the following Feynman graphs,
\begin{equation}
\begin{tabular}{ccc}
\parbox{18mm}{\includegraphics{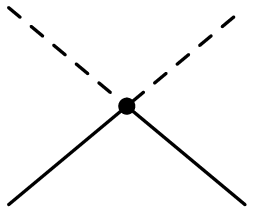}} $\sim  \alpha_{(4,a,2)} \frac{E^4}{v^2},$ &
\parbox{18mm}{\includegraphics{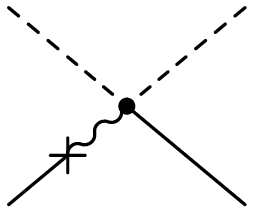}}$\sim  \alpha_{(4,a,2)} \frac{E^4}{v^2}\left(\frac{v}{M}\right)^2,$ &
\parbox{18mm}{\includegraphics{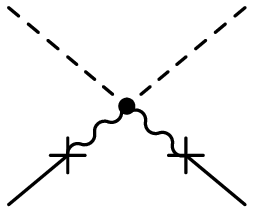}}$\sim  \alpha_{(4,a,2)} \frac{E^4}{v^2}\left(\frac{v}{M}\right)^4.$
\end{tabular}
\end{equation}
Hence, the contributions to this process coming from vertices obtained in the perturbative expansion of the covariant terms (\ref{dim6}) are governed by the leading $I_\gamma =0$ diagrams.

Radiative corrections to these coupling constants come from one-particle irreducible loop diagrams with the same number of external legs and the same power-law $E$ scaling (we saw in section \ref{running} that loop corrections with a different $E$ scaling correct coefficients of vertices containing a different number of derivatives). Hence, by (\ref{D}) these can only be one-loop diagrams constructed from the two-derivative Lagrangian. Using (\ref{type1}) we find that these diagrams have the size,
\begin{equation}\label{loop}
\mathcal{M}_{loop} \sim    \left(\frac{E}{m}\right)^{4} \left(\frac{m}{v}\right)^{4}\left(\frac{v}{M}\right)^{2I_\gamma}.
\end{equation}
Finding the largest loop correction therefore amounts to minimising the power of $v/M$ in (\ref{loop}). It is easily seen that without a direct aether-matter coupling constant at two-derivative order, a one-loop graph with both external $\pi$ and $\phi$ legs must contain at least two internal gravitons, $I_\gamma \geq 2$. Possible loop graphs with $I_\gamma < 2$ are either not connected and therefore do not contribute, or not one-particle irreducible. Hence the leading contribution is given by $I_\gamma =2$. Examples of these graphs include,
\begin{equation}
\begin{tabular}{ccc}
\parbox{20mm}{\includegraphics{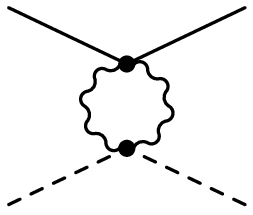}}$\,\sim \left(\frac{E}{M}\right)^{4}\,$ &
\parbox{20mm}{\includegraphics{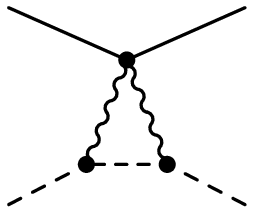}}$\sim \left(\frac{E}{M}\right)^{4}\,$ &
\parbox{20mm}{\includegraphics{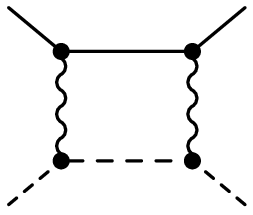}}$\quad\sim \left(\frac{E}{M}\right)^{4}\,$
\end{tabular}
\end{equation}

Hence to order $E^4$ we have the following contributions to the $E_\pi = E_\phi =2$, $E_\gamma = 0$ process,
\begin{equation}
M(E) \sim \frac{E^4}{v^2}\left[ \alpha_{(4,a,2)}(\mu)  + \frac{1}{M^2} \left(\frac{v}{M}\right)^{2} \log{\frac{E}{\mu}}\right]+ \mathcal{O}(E^6).
\end{equation}
Again, having subtracted any divergences (in $\overline{MS}$, for example) and then using the property (\ref{condition}), we see that under a change in the renormalisation point an order $\frac{v^2}{M^4}$ correction to $\alpha_{(4,a,2)}(\mu)$ is generated. Hence, as an NDA estimate of the size of these couplings we obtain,
\begin{equation}
\alpha_{(4,a,2)} \sim \left(\frac{v}{M}\right)^2 \frac{1}{M^2}, \qquad a=2,4.
\end{equation}
These couplings are therefore naturally smaller than the usual $(E/M)^2$ suppression by a factor of $(v/M)^2$. Hence if $v$ is sufficiently small compared to $M$, then $M \sim M_P$ and the dimension-six Lorentz violating $\phi$ operators are naturally small. An exact calculation of these corrections to a more realistic matter content could be used as a constraint on the size of $v/M$, given the experimental bounds found in ~\cite{Maccione:2009ju}, for example. This would subsequently provide an upper limit on the size of the usual phenomenological parameters,  $c_i \equiv (v/M)^2 \tilde{c_i}\sim (v/M)^2$. Tentatively assuming a similar bound for our idealised matter content we obtain~\footnote{There is a potential caveat here. If one invokes supersymmetry at higher energies in order to suppress any relevant and marginal Lorentz violating operators, then any $E^4$ contributions to the matter dispersion relation are naturally small~\cite{GrootNibbelink:2004za,Bolokhov:2005cj}. In this case the coupling constants for higher-dimension supersymmetric operators are naturally compatible with experimental constraints, and hence a weaker bound would be obtained.}, $c_i \lesssim 10^{-6}$. 

We will now comment on the use of a momentum-space cut-off regulator. Even if one were to overlook the requirement of symmetry preservation in the renormalisation procedure, there would still be the problem that this cut-off regulated theory has no controlled expansion. This is because there are potentially an infinite number of one-loop corrections from vertices containing arbitrarily many derivatives, all of which could contribute the same order of correction. In this set-up, in contrast with the dimensionally regulated case, the couplings in the two derivative Lagrangian could receive power-law corrections, the size of each individual correction dependent on the value of the cut-off chosen. For example, taking $\Lambda \sim v$ and using the NDA approach to estimate the size of the two-derivative direct coupling, one finds 
\begin{equation}\label{ultranaive}
\alpha_{(2,2,2)} \sim \left(\frac{v}{M}\right)^4.
\end{equation} 
Whilst it is interesting that this suggests the theory does not require finely-tuned cancellations if $v/M$ is small enough, it is inconclusive given that there is no controlled effective expansion.

\section{Discussion}
In this paper we considered the effective theory for quantised aether and metric perturbations of Einstein-aether. With the inclusion of scalar matter we estimated the size of a general Feynman diagram, and demonstrated that the theory can be truncated in a controlled way when a dimensional regulator is used. We showed that the phenomenological $c_i$ parameters of Einstein-aether theory receive only negligible logarithmic corrections, justifying the scale independence which is assumed when they are compared to experiment. We estimated the expected size of Lorentz violation for matter due to loop effects communicating the Lorentz violation between the aether and matter sectors, and outlined how this could be used as a new constraint on the $c_i$ parameters.

With a controlled way to truncate this effective theory it becomes possible for more detailed features to be investigated. It would be interesting to study the inclusion of non-relativistic masses into this low-energy theory, and to see whether a good effective expansion is retained. This has been investigated in the case of general relativity in \cite{Donoghue:1996mt}, where a good expansion is found. Furthermore, the leading quantum corrections to the Newtonian potential has been calculated for pure gravity~\cite{Donoghue:1993eb,Donoghue:1994dn}. It would be interesting to see how the aether perturbations modify this result, and its dependence on the observer frame in which the perturbations were defined.

There are of course a number of technical issues which have not been addressed here. A step towards a background field method for this theory may be found in the context of chiral perturbation theory~\cite{Gasser:1983yg}, which provides a way to deal with a similar type of length-fixing constraint in a covariant way. In pure gravity infrared divergences can be cancelled by graphs with additional soft gravitons~\cite{Weinberg:1965nx}, something which has not been investigated here. Additionally, perturbation theory in pure gravity is renormalisable in the effective field theory sense~\cite{Gomis:1995jp}; for Einstein-aether we have  assumed this property.

\acknowledgments
I would like to thank Toby Wiseman for suggesting this project, and for helpful discussions. I would also like to thank Christopher Eling, Brendan Z. Foster and David Mattingly for useful comments, and Arttu Rajantie for enlightening conversations. This work was supported by an STFC studentship.

\providecommand{\href}[2]{#2}\begingroup\raggedright\endgroup

\end{document}